\documentclass{emulateapj}
\usepackage[dvips]{color}
\usepackage{graphicx}
\usepackage{subfloat}
\def\msol{\hbox{\kern 0.20em $M_\odot$}}
\def\lsol{\hbox{\kern 0.20em $L_\odot$}}
\def\rsol{\hbox{\kern 0.20em $R_\odot$}}
\def\sr{\hbox{\kern 0.20em sr}}
\def\srmu{\hbox{\kern 0.20em sr$^{-1}$}}
 
\def\g{\hbox{\kern 0.20em g}}
\def\gmu{\hbox{\kern 0.20em g$^{-1}$}}
\def\kg{\hbox{\kern 0.20em kg}}
\def\pc{\hbox{\kern 0.20em pc}}
 
\def\mum{\hbox{\kern 0.20em $\mu$m}}
\def\mumd{\hbox{\kern 0.20em $\mu$m$^{-2}$}}
\def\cm{\hbox{\kern 0.20em cm}}
\def\m{\hbox{\kern 0.20em m}}
\def\km{\hbox{\kern 0.20em km}}
\def\nm{\hbox{\kern 0.20em nm}}
 
\def\s{\hbox{\kern 0.20em s}}
\def\h{\hbox{\kern 0.20em h}}
\def\sec{\hbox{\kern 0.20em sec}}
\def\min{\hbox {\kern 0.20em min}}
\def\smu{\hbox{\kern 0.20em s$^{-1}$}}
\def\smd{\hbox{\kern 0.20em s$^{-2}$}}
\def\an{\hbox{\kern 0.20em an}}
\def\anmu{\hbox{\kern 0.20em an$^{-1}$}}
\def\deg{\hbox{\kern 0.20em $^{\rm o}$}}
\def\yr{\hbox{\kern 0.20em yr}}
\def\yrmu{\hbox{\kern 0.20em yr$^{-1}$}}
\def\Myr{\hbox{\kern 0.20em Myr}}
\def\Mymu{\hbox{\kern 0.20em Myr$^{-1}$}}
\def\K{\hbox{\kern 0.20em K}}
\def\pcmu{\hbox{\kern 0.20em pc$^{-1}$}}
\def\pcmd{\hbox{\kern 0.20em pc$^{-2}$}}
\def\pcmt{\hbox{\kern 0.20em pc$^{-3}$}}
\def\kms{\hbox{\kern 0.20em km\kern 0.20em s$^{-1}$}}
\def\kmpd{\hbox{\kern 0.20em km$^{2}$}}
\def\kpc{\hbox{\kern 0.20em kpc}}
\def\cms{\hbox{\kern 0.20em cm\kern 0.20em s$^{-1}$}}
\def\erg{\hbox{\kern 0.20em erg}}
\def\ergs{\hbox{\kern 0.20em erg}}
\def\cmpd{\hbox{\kern 0.20em cm$^2$}}
\def\cmmd{\hbox{\kern 0.20em cm$^{-2}$}}
\def\cmms{\hbox{\kern 0.20em cm$^{-6}$}}
\def\cmpt{\hbox{\kern 0.20em cm$^3$}}
\def\cmmt{\hbox{\kern 0.20em cm$^{-3}$}}
\def\mpd{\hbox{\kern 0.20em m$^2$}}
\def\mmd{\hbox{\kern 0.20em m$^{-2}$}}
\def\mpt{\hbox{\kern 0.20em m$^3$}}
\def\mmt{\hbox{\kern 0.20em m$^{-3}$}}
\def\mujy{\hbox{\kern 0.20em $\mu$Jy}}
\def\mjy{\hbox{\kern 0.20em mJy}}
\def\Mj{\hbox{\kern 0.20em MJy}}
\def\jy{\hbox{\kern 0.20em Jy}}
\def\ghz{\hbox{\kern 0.20em GHz}}
\def\srmd{\hbox{\kern 0.20em sr$^{-1}$}}
\def \kms{km~$\rm{s}^{-1}$}

\def \mum{$\mu$m}
\def\G{\hbox{\kern 0.20em G}}

\slugcomment{Submitted 27Jan06, Accepted 8Apr06, to appear in ApJ}

\shorttitle{Cold Dust in NGC205}
\shortauthors{Marleau et al.}

\begin{document}

 \title{Mapping and Mass Measurement of the Cold Dust in NGC~205
 with Spitzer}

\author{F. R. Marleau\altaffilmark{1},
A. Noriega-Crespo\altaffilmark{1},
K.A. Misselt\altaffilmark{2},
K.D. Gordon\altaffilmark{2},
C.W. Engelbracht\altaffilmark{2},
G.H. Rieke\altaffilmark{2},
P. Barmby\altaffilmark{3},
S.P. Willner\altaffilmark{3},
J. Mould\altaffilmark{4},
R.D. Gehrz\altaffilmark{5},
and
C.E. Woodward\altaffilmark{5}}

\altaffiltext{1}{SPITZER Science Center, California Institute of 
Technology,
CA 91125  USA}
\altaffiltext{2}{Steward Observatory, University of Arizona, Tucson AZ 
85721}
\altaffiltext{3}{Harvard-Smithsonian Center for Astrophysics, 60 Garden
Street, Cambridge, MA 02138}
\altaffiltext{4}{NOAO, P.O. Box 26732, Tucson, AZ 85726}
\altaffiltext{5}{Astronomy Department, University of Minnesota, Minneapolis, 
MN 55455}

\begin{abstract}
We present observations at 3.6, 4.5, 5.8, 8, 24, 70 \& 160\mum\ of
NGC~205, the dwarf elliptical companion of M31, obtained with the {\it
Spitzer Space Telescope}. The point sources subtracted images at 8 and
24\mum\ display a complex and fragmented infrared emission coming from
both very small dust particles and larger grains.  The extended dust
emission is spatially concentrated in three main emission regions,
seen at all wavelengths from 8 to 160\mum. These regions lie
approximately along NGC~205's semi-major axis and range from $\sim$100
to 300~pc in size. Based on our mid-to-far infrared flux density
measurements alone, we derive a total dust mass estimate of the order
of $3.2 \times 10^4 M_\odot$, mainly at a temperature of
$\sim$~20K. The gas mass associated with this component matches the
predicted mass returned by the dying stars from the last burst of star
formation in NGC~205 ($\sim$0.5~Gyr ago). Analysis of the {\it
Spitzer} data combined with previous 1.1mm observations over a small
central region or ``Core'' (18\arcsec~diameter), suggest the presence
of very cold (T$\sim$ 12K) dust and a dust mass about sixteen times
higher than is estimated from the {\it Spitzer} measurements
alone. Assuming a gas to dust mass ratio of 100, these two datasets,
i.e.\ with and without the millimeter observations, suggest a total
gas mass range of $3.2 \times 10^6$ to $5 \times 10^7 M_\odot$.
\end{abstract}

\keywords{galaxies: dwarf --- galaxies: individual (NGC~205) 
--- galaxies: ISM --- infrared: galaxies --- infrared: ISM 
--- ISM: dust, extinction  --- ISM: structure}

\lefthead{Marleau et al.}

\righthead{FIR Mapping of NGC205}

\section{Introduction}

The galaxies of the Local Group, because of their proximity and our
ability to carry out high spatial resolution studies in nearby
systems, provide us with powerful templates to understand the
chemical, morphological and kinematical characteristics, plus star
formation history, of more distant galactic systems. The Local dwarf
galaxies (LDGs), in particular, have been subject to numerous recent
studies of their stellar populations and star formation (for summaries
see for example, Hodge 1989, Mateo 1998, Grebel 2005).  Dwarf
ellipticals follow a different ``fundamental plane'' than true
ellipticals (Ferguson \& Binggeli 1994), nevertheless they do share
many common characteristics, in particular the presence of a mostly
old stellar population (Mateo 1998). One of these local dwarf systems
is NGC~205, a low surface brightness dwarf elliptical companion of
M31. NGC~205 is interesting because of its very conspicuous dark
clouds, which were detected in some of the earliest photographic
plates of M31 and its surroundings (Baade 1944; Hodge 1973)(NGC~205
lies $\sim 36\arcmin$ NW from the M31 nucleus).  These visually dark
clouds reveal the presence of an interstellar medium in a stellar
population dominated by old stars, that is, in an elliptical galaxy.

There are several indicators for an intermediate and young stellar
population, as well as for gas enrichment, in NGC~205.  A handful of
blue and UV-bright stars have been detected in its nucleus (Baade
1951; Hodge 1973; Bertola et al. 1995), with ages ranging from $\sim 5
\times 10^8$ yr to a few $10^6$ yr. In terms of metallicity, the
location of its giant branch implies a mean value of ${\rm [Fe/H]} =
-0.85\pm0.2$ (Mould, Kristian \& Da Costa 1984), higher than for most
LDGs, which have values ranging from $-1.1$ to $-2.2$ (Mateo
1998). Moreover, a stellar population study of the resolved bright
near-IR stars in NGC~205 (Davidge 2003) indicates that the brightest
asymptotic giant branch stars formed a few tenths of a Gyr ago. All
these indicators suggest an episode of star formation in NGC~205,
which took place a few $10^8$ years ago, and a recent replenishment of
the interstellar material by this population of evolved
stars. Although there is no consensus on the process which triggered
such a burst, gravitational tidal interaction of NGC~205 with M31 has
always been an appealing mechanism because it simultaneously explains
the twisted surface brightness isophotes (Hodge 1973; Choi et
al. 2002), the tidal trail of blue metal poor RGB stars (McConnachie
et al. 2004; Ibata et al. 2001) and the kinematically distinct
behavior and morphology of its interstellar medium, particularly of
the HI gas (Young \& Lo 1997).

The ISM in NGC~205 has some peculiarities. Based on the last burst of
star formation $\sim 5 \times 10^8$ yr ago (Wilcots et al. 1990;
Bertola et al.  1995), one can estimate the amount gas injected back
into the ISM, which is $\sim 2.6\times 10^6 M_{\odot}$ (following
Faber \& Gallagher 1976). This mass is about three times larger than
that measured by Sage, Welch \& Mitchell (1998) of $7.2\times10^5
M_{\odot}$ (from CO (H$_2$) plus HI, scaled by a factor 1.4 to include
Helium). This difference in masses led to the idea that gas is being
removed by the young stars in NGC~205, either by stellar winds or
supernova explosions (Welch, Sage \& Mitchell 1998). Recently Haas
(1998) carried out photometric observations with the {\it Infrared
Space Observatory (ISO)} between 120 and 200\mum\ that, combined with
a millimetric measurement of the central 18\arcsec\ (Fich \& Hodge
1991), yielded large amounts of cold dust with mass $0.2-1.2 \times
10^4 M_\odot$ at $T\sim 10$~K. Assuming a gas-to-dust mass ratio of
100, the cold dust traces nearly $10^6$ M$_\odot$ of gas at
center. There is vast observational evidence for ``warm'' and ``cool''
dust in elliptical galaxies based on mid/far infrared data (see
e.g. Jura et al. 1987; Knapp et al. 1989; Ferrari et al. 2002; Temi et
al. 2004; Xilouris et al. 2004). The initial IRAS measurements were
limited to detection of temperatures of $\sim 30$ K traceable by the
100\mum\ fluxes. This led to the possibility of missing a colder
component with nearly 90\% of the dust mass at lower temperatures
(T~$\sim 15-20$K) which is radiating at longer wavelengths ($>$
100\mum) (see e.g. Devereux \& Young 1990).  The longer wavelength far
infrared observations ($>$ 100\mum) combined with millimeter and
sub-millimeter measurements have confirmed this picture in several
early-type galaxies, where the bulk of the dust mass is cold ($\sim
20$ K) and nearly an order of magnitude larger than previously
estimated (Wiklind \& Henkel 1995; Temi et al. 2004).

This study presents new mid/far infrared images of NGC~205 obtained
with the {\it Spitzer Space Telescope (Spitzer)}. The higher
sensitivity and spatial resolution enable us for the first time to
look {\it directly} into the detailed structure and properties of the
dust clouds in NGC~205 and to evaluate their dust properties and
content. The outline of the paper is as follows. In Section~2, we
describe the {\it Spitzer} observations and the data reduction. The
{\it Spitzer} images are presented in Section~3, along with a
description of the mid/far-IR emission morphology and a comparison
with other tracers of the ISM. Additional image processing is
discussed in Section~4. Dust mass measurements and gas mass estimates
are derived in Section~5 using two different methods. In Section~6, we
discuss the results with some of the previous mass estimates and
theoretical expectations. We conclude with a summary of our findings
in Section~7.

\section{Observations}

The mid-IR observations were obtained with IRAC, the infrared camera
on board {\it Spitzer} (Fazio et al. 2004) on 20 January 2005 as part
of the GO1 program to map M31 (PI: Barmby, Program ID: 3126). The
region around NGC~205 was observed with four dithered 30 sec frames
per sky position; the IRAC field of view (FOV) is 5\arcmin. The images
obtained with channels 1 and 2 (at 3.6 and 4.5\mum) are used mostly to
determine the stellar contribution to the overall spectral energy
distribution of NGC~205.  The channels 3 and 4 (at 5.8 and 8\mum)
contain the most information about the dust properties; channel 4 in
particular, with a passband centered at 7.9\mum\ includes some of the
strongest Polycyclic Aromatic Hydrocarbons (PAH) emission features
known (e.g. bands at 7.7 and 8.6\mum); these are considered some of
the best tracers of very small particles (see e.g. Li \& Draine 2001,
Draine \& Li 2001 and references therein). The Basic Calibrated Data
produced by the SSC pipeline (Version 11) were combined with the SSC
MOPEX software to produce final mosaics with pixel size 1.2\arcsec
(all channels) and spatial resolutions of less than 2\arcsec\ FWHM.

\begin{figure}
\centerline{
\includegraphics[width=260pt,height=260pt,angle=0]{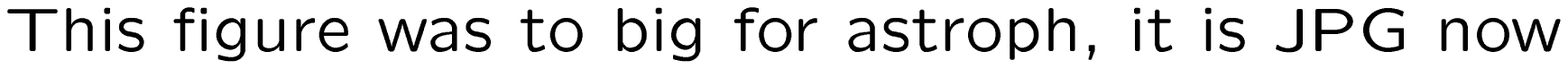}}
\caption{\label{fig:1} IRAC 3.6, 4.5, 5.8 \& 8\mum\ images of
NGC~205. The FOV of each image is 15\arcmin\ $\times$ 15\arcmin\ with N
up and E to the left.}
\end{figure}

\begin{figure}
\centerline{
\includegraphics[width=260pt,height=260pt,angle=0]{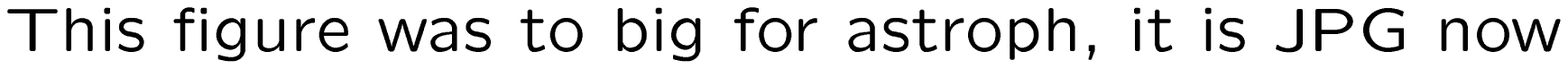}
}
\caption{\label{fig:2} IRAC 8\mum\, MIPS 24, 70 \& 160\mum\ images of
NGC~205. The FOV of each image is 5\arcmin\ $\times$ 5\arcmin\ with N
up and E to the left.}
\end{figure}

The far-IR data were obtained with the multiwavelength photometer MIPS
onboard {\it Spitzer} on 25 August 2004 (Program ID 99). These
observations were taken as part of the large M31 mosaic consisting of
7 scan legs, between 0.75 and 1.25 degree long, with a 148\arcsec\
overlap between scan legs. The scanning was done at the medium scan
rate covering a total area of 1\arcdeg\ $\times$ 3\arcdeg\,
approximately oriented along the major axis of M31, with a total
observing time of 17.1 hrs. The NGC~205 observations themselves have a
depth of 84 sec per pixel. The details of the observations and
reductions are given by Gordon et al. (2006). In the three MIPS bands
a background subtraction has been performed, that removes some time
dependent artifacts along the scan legs. In the 70 and 160\mum\ images
some stripping remains due to a non-linear time response of the Ge:Ga
arrays, however their overall effect on the flux density around
NGC~205 is very small ($\sim 1-2$\%), and this is included in our
uncertainties (see Section~4.1). The final 24, 70, and 160 micron
images have spatial resolutions of 6, 18 and 40\arcsec\ FWHM,
respectively (Rieke et al. 2004). The mosaic pixel scales are
2.49\arcsec\ at 24\mum, 9.85\arcsec\ at 70\mum, and 16.0\arcsec\ at
160\mum.

The {\it Spitzer} data are complemented with ISOCAM observations of
NGC~205 from the ISO archive. The archive data (Off Line Processing
V10.0) was re-reduced using the CIA (Cam Interactive Analysis V4.0)
software to improve mainly on the deglitching, following the rest of
the standard reduction steps (Coulais \& Abergel 2000; Blommaert et
al.  2003). The data set includes images at 14.3\mum\ covering a
204\arcsec $\times$ 216\arcsec\ field of view and with 6\arcsec\ per
pixel. All the measurements were performed with background
subtraction, and were compared with the global estimates of Xilouris
et al. (2004) for consistency.

For both IRAC and MIPS we used a 15\arcmin\ $\times$ 15\arcmin\
section of the M31 maps centered on the position of NGC~205
[$\alpha(2000)$ = 00$^h$40$^m$22.1$^s$, $\delta(2000)$ = 41\arcdeg
41\arcmin 07.1\arcsec]. This region is large enough to encompass most
of visual size of the galaxy of 9\arcmin\ $\times$ 19\arcmin\ (25
B~mag arcsec$^{-2}$) (Nilson 1973), and certainly all the mid/far IR
emission (Rice et al. 1988). We adopt a distance to NGC~205 of
$815\pm35$~kpc (Mateo 1998; McConnachie et al. 2004) throughout this
work, i.e. at that distance, spatial structures 1\arcsec\ across
measure $\sim$ 3.9~pc. This tentatively puts NGC~205 behind M31, which
is at a distance of 780~kpc (Rich et al. 2005; McConnachie et al. 2005).

\begin{figure}
\centerline{
\includegraphics[width=210pt,height=210pt,angle=0]{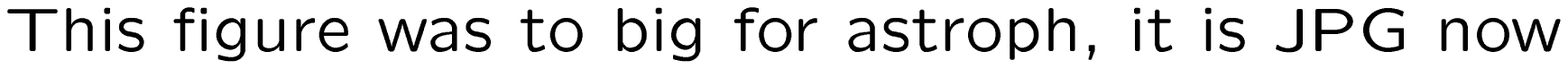}}
\caption{\label{fig:3} A three color image of NGC~205 as seen by
IRAC at 3.6 (blue), 5.8 (green) and 8\mum\ (red). The image emphasizes
the dust clouds distribution seen at longer wavelengths over a 15\arcmin\
square region, with N up and E to the left.}
\end{figure}

\section{Morphology of Infrared Emission}

Figure~1 and 2 show the images of NGC~205 taken with IRAC at 3.6 \&
4.5\mum\ (stellar emission dominant, see Table~4) and at 5.8, 8, 24,
70 \& 160\mum\ (dust emission dominant, see Table~4). Figure~3
displays a three color image of NGC~205 where the dust clouds at 5.8
and 8\mum\ stand out within the stellar background detected by the
shorter wavelength bands. The higher angular resolution images at 8
and 24\mum\ display a spatially complex and fragmented extended
emission. We expect that the main contribution to the 8\mum\ emission
arises from aromatic molecules and at 24\mum\ from submicron-sized
grains (Li \& Draine 2002; Xilouris et al. 2004). The overlay of these
two wavelengths (Fig.~4) shows the close spatial relationship between
these two dust distributions. However, the 8\mum\ image shows more
extended emission than at 24\mum\ (see Fig.~5).

\begin{figure}
\centerline{
\includegraphics[width=210pt,height=210pt,angle=0]{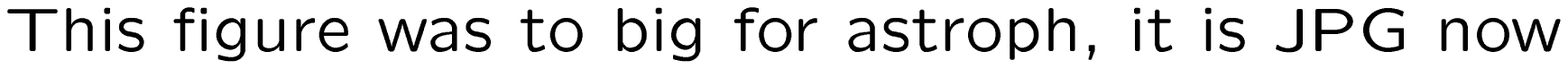}
}
\caption{\label{fig:4} The three main emission regions -- ``North'',
``Center'', ``South''-- as well as the ``Core'' and ``Total'' regions
(as defined in Table~1) are shown overlayed on the IRAC 8\mum\ image
of NGC~205. The FOV is 5\arcmin\ $\times$ 5\arcmin\ with N up and E to
the left.}
\end{figure}

The morphology of the diffuse emission is complex and clumpy;
nevertheless we can clearly identify three spatially distinct emission
regions: the largest region to the south with a size of $\sim$
76\arcsec\ $\times$ 37\arcsec\ ($\sim$ 300~pc $\times$ 146~pc), the
central region with a size of $\sim$ 49\arcsec\ $\times$ 28\arcsec\
($\sim$ 193~pc $\times$ 110~pc), and the smallest region to the north
with a size $\sim$ 33\arcsec\ $\times$ 31\arcsec\ ($\sim$ 130~pc
$\times$ 123~pc) (see Fig.~4 and Table~1). Even at the lower
resolution of the 70 and 160\mum\ bands (see Fig.~2), where the bulk
of the emission is produced by dust emission from the larger grains
($\sim$ 0.1-0.5\mum, D\'esert, Boulanger \& Puget 1990; Li \& Draine
2001), one can easily identify the same three large regions seen at
shorter wavelengths. We have included also a ``Core'' region with a
diameter of 18.\arcsec4 for which 1.1mm observations are available
(Fich \& Hodge 1991).

\begin{table}[h!]
\centerline{TABLE 1}
\centerline{NGC~205 Emission Regions \label{tbl-1}}
\vspace{0.2cm}
\begin{center}
\begin{tabular}{cccc}
\tableline\tableline
Region\tablenotemark{a}        &RA (J2000)   &Dec (J2000)  &Area \\
            & $^h$~~$^m$~~$^s$ & \arcdeg~~~ \arcmin~~~ \arcsec\ &\arcmin$^2$ \\
\tableline
North                          &0 40 24.00  &41 41 49.9   &0.24 \\
Center                         &0 40 22.29  &41 41 09.4   &0.30 \\
South                          &0 40 23.75  &41 40 10.3   &0.55 \\
Core                           &0 40 22.29  &41 41 09.4   &0.07 \\
Total                          &0 40 22.38  &41 40 52.3   &4.36 \\
\tableline
\end{tabular}
\tablenotetext{a}{elliptical aperture as shown in Fig.~4}
\end{center}
\end{table}

The 14.3\mum\ ISOCAM observations, however, cover a smaller FOV and
the final mosaic has some vignetting that leaves an effective FOV of
$\sim 192$\arcsec\ radius. Although our mosaic is a bit larger than
that presented by Xilouris et al. 2004 (Fig.~A.1), the `South'' and
``Total'' regions are 3\% and 6\% smaller at 14.3\mum, respectively,
than at all other wavelengths (see Table~3). Therefore, the emission
not included directly in these two regions, given the very low surface
brightness at 14.3\mum, is small.

\begin{figure}
\centerline{
\includegraphics[width=210pt,height=210pt,angle=0]{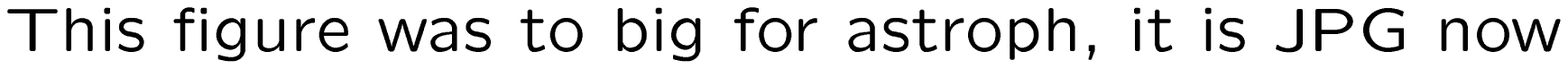}
}
\caption{\label{fig:5} IRAC 8\mum\ grayscale image of NGC~205 where point
sources have been removed to reveal more clearly the extended diffuse
emission. The image is displayed at high contrast to enhance the faint
emission surrounding the bright central region. The FOV is 15\arcmin\
$\times$ 15\arcmin\ with N up and E to the left.}
\end{figure}

A comparison of the mid/far IR data with other tracers of the ISM,
such as HI (Young \& Lo 1997), shows the same morphology (see Fig.~6),
indicating that dust and gas are very well mixed in NGC~205.  Based on
the HI velocity data, it is clear that the HI gas rotates in the
North-South direction with a velocity gradient of 42~km~s$^{-1}$ over
900~pc (Young \& Lo 1997). However, there is no clear one-to-one
correspondence between the HI column density peaks and velocity, and
therefore, the three distinct IR emission regions are not necessarily
dynamical entities. The observed dust morphology is similar at all
wavelengths from 8 to 160\mum\ (within the limits of varying angular
resolution) and also similar to the HI morphology seen in the 21-cm
line. The data is consistent with an ISM in which the atomic gas and
dust are well mixed and therefore we see no evidence for significant
stratification of the ISM in NGC 205.

\begin{figure}
\centerline{
\includegraphics[width=210pt,height=210pt,angle=0]{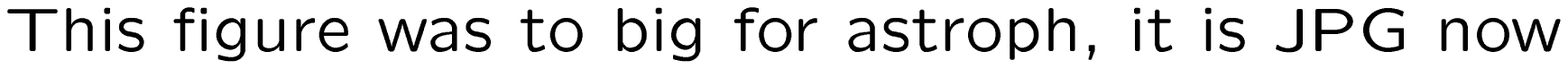}
}
\caption{\label{fig:6} The HI column density derived by Young \& Lo
(1997) are shown overlayed on the IRAC 8\mum\ grayscale image of
NGC~205 where point sources have been removed. The HI density
contours range from 2 $\times 10^{19}$ cm$^{-2}$ to 3.8 $\times
10^{20}$ cm$^{-2}$ in intervals of 4 $\times 10^{19}$ cm$^{-2}$.
The FOV is 5\arcmin\ $\times$ 5\arcmin\ with N up and E to the left.}
\end{figure}

\section{Additional Data Processing}

A reliable estimate of the IR and diffuse emission in NGC~205 requires
an estimate of the foreground point source contamination arising
either from the Milky Way (MW) or M31.  Point sources were extracted
in the central 5\arcmin\ $\times$ 5\arcmin\ region of NGC~205 in the
5.8, 8.0 and 24\mum\ images using the source extraction software
StarFinder (Diolaiti et al. 2000).  {\it Spitzer} [8]-[24] and
[5.8]-[8] colors were derived for these point sources and compared to
the colors of a sample of Galactic AGB and K Giant stars templates in
the mid-infrared observed with the Short Wavelength Spectrometer (SWS)
on board of ISO. Using ISAP (ISO Spectral Analysis Package Version
2.0)\footnote{ISAP is available at
http://www.ipac.caltech.edu/iso/isap/isap.html} it is possible to
determine the flux density within each of the Spitzer's
imager/photometer bandpass over the SWS spectra which cover a
wavelength range of 2.4-45.2\mum. The point sources in the central
region of NGC~205 (Fig.~7 top, open squares) have colors similar to
those of AGB stars (filled triangles). However, K Giant stars in the
field are clearly identifiable due to their much ``bluer'' [8]-[24]
color and, consequently, we conclude that none are present in the
central region of NGC~205.

The foreground contamination at 24\mum, from M31 point sources, where
dusty AGBs contribute most of their flux, is estimated using six
control fields. Each field has dimension 15\arcmin\ $\times$
15\arcmin\ to match that of NGC~205 and is located a comparable
distance away from the center of M31 (Table~2).  To determine the
number of AGB stars in NGC~205, we count these stars in the central
15\arcmin\ square region and estimate the number of foreground stars
by averaging their numbers in the six control fields. Based on the
results shown at the bottom of Figure~7, we estimate that about half
of the point sources in the NGC~205 15\arcmin\ $\times$ 15\arcmin\
field are from foreground contamination due either to the MW or M31.

We estimate that in the smaller ``Total'' region centered on NGC~205
(as defined in Table~1), about a third of the point sources are from
foreground contamination. Given a mean flux density of 0.6 mJy for
contaminating point sources (see Fig.~7, bottom), we estimate a flux
density contamination of $\sim$~2 mJy. This number is below our flux
density measurement errors at 24\mum\ (see Table~3) and therefore no
further processing, i.e. point source removal, is done to the 24\mum\
image.

\begin{figure}
\centerline{\includegraphics[width=260pt,height=200pt,angle=0]{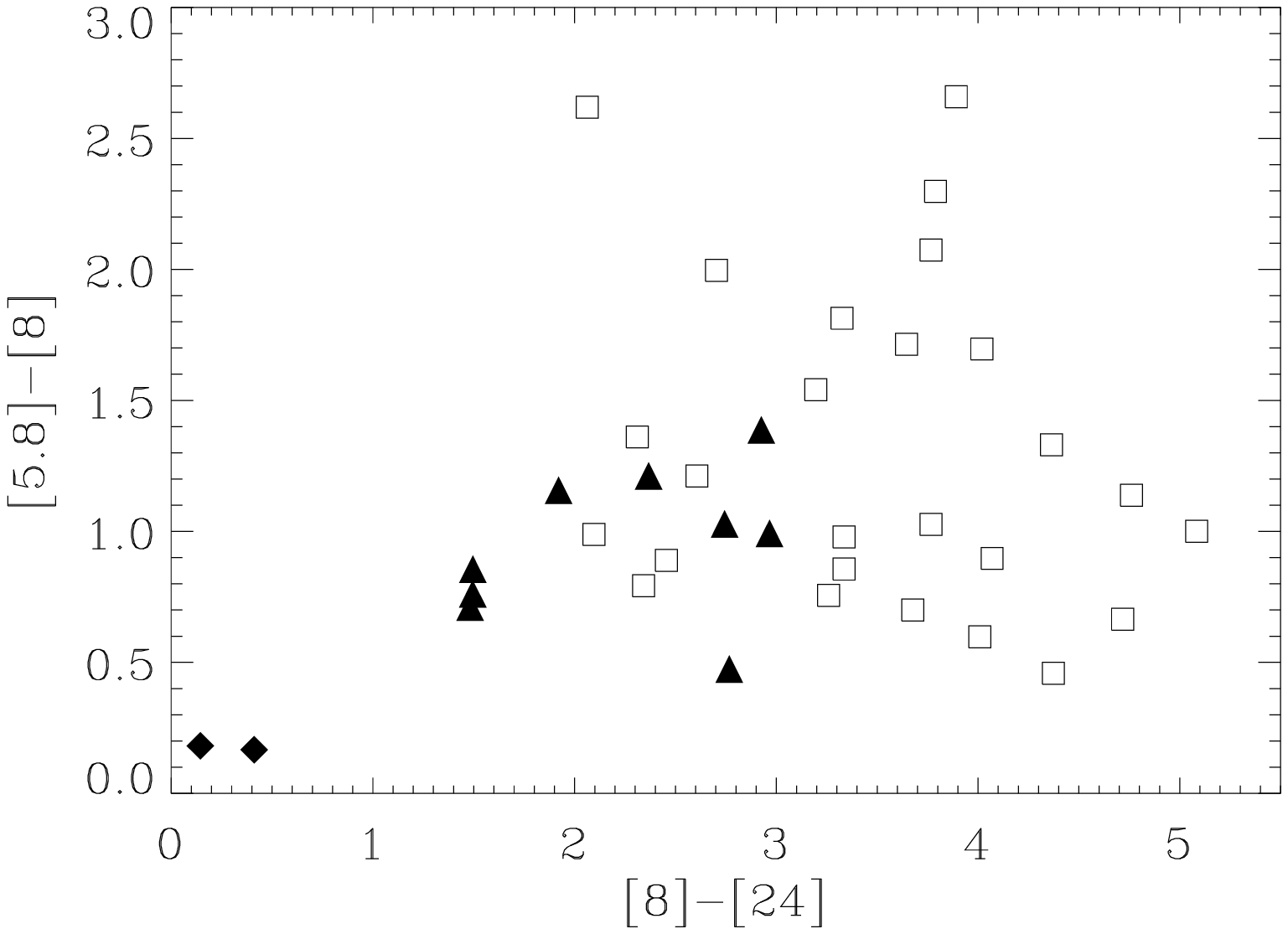}}
\centerline{\includegraphics[width=260pt,height=200pt,angle=0]{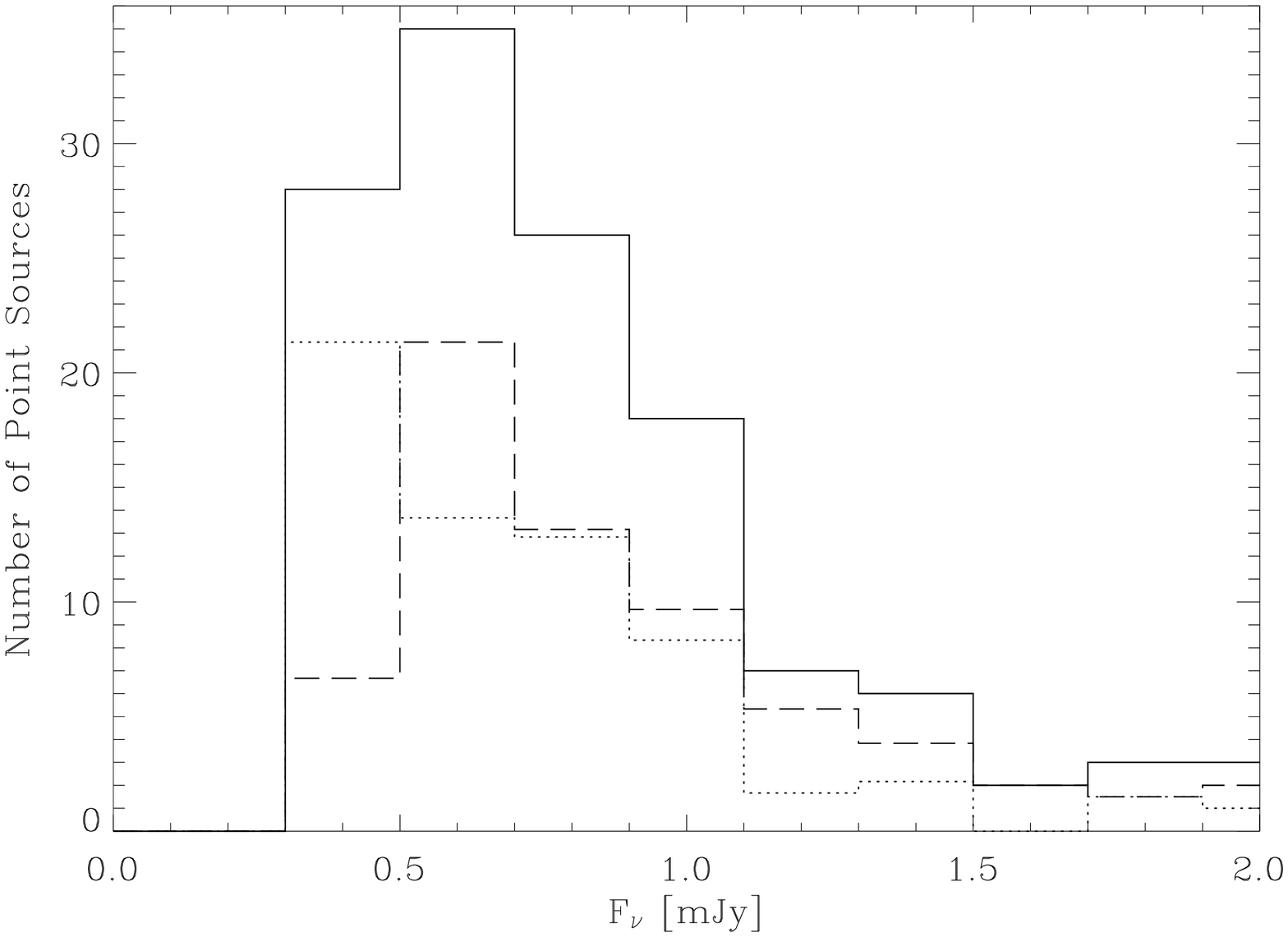}}
\caption{\label{fig:7} {\it Top:} Color-color diagram of a sample of
K Giant stars ({\it filled diamonds}),  AGB stellar templates ({\it
filled triangles}), and the point sources extracted in the central
5\arcmin\ of the NGC~205 field ({\it open squares}). {\it Bottom:} Histogram
showing the number of point sources as a function of 24\mum\ flux in
the central 5\arcmin\ of the NGC~205 field ({\it solid  line}),
for the average of the six control fields (see Table~2; {\it dashed
line}), and the remaining distribution for NGC~205 after the
average from the control fields has been removed ({\it dotted line}).}
\end{figure}

\begin{table}[h!]
\centerline{TABLE 2}
\centerline{NGC~205 Pointing and Control Fields \label{tbl-2}}
\vspace{0.2cm}
\begin{center}
\begin{tabular}{ccc}
\tableline\tableline
Field        &RA (J2000)   &Dec (J2000) \\
            & $^h$~~$^m$~~$^s$ & ~\arcdeg~~ \arcmin~~~ \arcsec\ \\
\tableline
NGC 205          &0 40 22.10    &41 41 07.1 \\
control field a  &0 36 38.65    &40 37 47.0 \\
control field b  &0 44 54.11    &40 51 03.5 \\
control field c  &0 43 36.11    &40 31 49.4 \\
control field d  &0 46 24.97    &41 10 13.8 \\
control field e  &0 47 38.73    &41 28 52.5 \\
control field f  &0 42 54.21    &42 19 51.6 \\
\tableline
\end{tabular}
\end{center}
\end{table}

\subsection{Aperture Correction of Extended Sources}

We measure the IRAC 3.6, 4.5, 5.8, \& 8\mum\ and the MIPS 24, 70, \&
160\mum\ fluxes in the spatially resolved regions common at all
wavelengths, as depicted in Figure~4 and listed in Table~1.  The IRAC
and MIPS data are combined with previous observations taken with ISO
at 14.3\mum\ (Xilouris et al. 2004) and JCMT at 1.1mm (Fich \& Hodge
1991). The flux uncertainty for each region includes the combined
effect of the systematic errors due to the post processing of the
images, plus that of the flux density calibration of the instruments
(Fazio et al. 2004; Rieke et al. 2004). The flux uncertainty,
therefore, was estimated by measuring the residual flux density away
from the galaxy in the background subtracted images and adding an
absolute flux calibration uncertainty of 10\% for the IRAC and MIPS
24\mum\ measurements, and of 20\% for the ISO 14.3\mum\ and MIPS 70 \&
160\mum\ fluxes. The 8\mum\ fluxes are corrected (multiplied) by the
effective aperture correction factor of 0.944 (3.6\mum), 0.937
(4.5\mum), 0.772 (5.8\mum), and 0.737 (8\mum) (Reach et al. 2005). The
corrected flux densities and uncertainties are given in Table~3.

\begin{table}[h!]
\centerline{TABLE 3}
\centerline{Photometry of Emission Regions in NGC~205 \label{tbl-3}}
\vspace{0.2cm}
\footnotesize
\begin{center}
\begin{tabular}{cccccc}
\tableline\tableline
Wavelength & North & Center & South & Core & Total \\
\mum     & \multicolumn{5}{c}{Flux Density in mJy}\\
\tableline
1.22 & 430$\pm$2  & 130$\pm$7  & 110$\pm$6   &46.2$\pm$2 & 940$\pm$47 \\
1.65 & 550$\pm$3  & 160$\pm$8  & 40$\pm$7    &56.8$\pm$3 & 116$\pm$58 \\
2.16 & 380$\pm$2  & 110$\pm$6  & 100$\pm$5   &39.1$\pm$2 & 820$\pm$41 \\
3.6  & 21$\pm$3   & 59$\pm$6   & 54$\pm$7    &23.5$\pm$2 & 438$\pm$52 \\
4.5  & 13$\pm$2   & 35$\pm$4   & 33$\pm$5    &14.9$\pm$1 & 266$\pm$33 \\
5.8  & 15$\pm$3   & 33$\pm$5   & 39$\pm$9    &14.2$\pm$1 & 279$\pm$62 \\
8    & 12$\pm$1   & 26$\pm$3   & 26$\pm$3    &14.5$\pm$1 & 148$\pm$15 \\
14.3 &  7$\pm$2   & 15$\pm$3   & 25$\pm$5    & 5.9$\pm$1 & 101$\pm$23 \\
24   & 10$\pm$4   & 18$\pm$5   & 26$\pm$10   & 5.8$\pm$1 & 141$\pm$23 \\
70   & 104$\pm$21 & 181$\pm$36 & 301$\pm$60  &61.7$\pm$12& 1428$\pm$290 \\
160  & 473$\pm$279 & 713$\pm$363&1490$\pm$692&253$\pm$45& 8761$\pm$4720 \\
\tableline
\end{tabular}
\end{center}
\end{table}

\subsection{Stellar Photospheric Component of Diffuse Emission}

The stellar photospheric component of the diffuse emission coming from
NGC~205 itself must be removed before the analysis of the dust
component. We conservatively assume that, in addition to the 2MASS J,
H, and K bands, all the light in the 3.6 and 4.5 IRAC bands is
photospheric in origin. The J, H, K, IRAC 3.6 and 4.5 filter response
functions were multiplied with a PEGASE stellar model (Fioc \&
Rocca-Volmerange 1997) with an age of 350~Myr (see Fig.~9) and
integrated over wavelength yielding fluxes that can be compared with
the observed fluxes. The age was picked to roughly coincide with the
(uncertain) epoch of recent star formation in NGC~205 (see eg. Butler
\& Martinez-Delgado 2005; Davidge 2005).  There is no diagnostic power
in the age assumed; we used synthetic spectra from Simple Stellar
Population (SSP) models only to estimate the stellar contribution to
the Spitzer bands, as simple power laws cannot reproduce the J, H, K
colors. Scalings were derived for each band individually and averaged
together to get the final scaling for the photospheric component. In
all cases, the J through 4.5\mum\ fluxes are consistent with being
completely stellar in origin. To estimate the photospheric
contribution to the flux density at longer wavelengths, we multiplied
and integrated over wavelength the relevant IRAC and MIPS bandpasses
with the PEGASE model and subtracted the result from the observed flux
density to yield an estimate for the total dust flux density. For the
relative contributions of the diffuse photospheric emission to the
observed fluxes, see Table~4. The age we use for the PEGASE models
(within a factor of 3) does not significantly alter the dust fits;
indeed, PEGASE models between 0.1$-$1 Gyr result in very similar
estimates for the stellar contamination in the Spitzer bands,
especially for $\lambda \ge 8$\mum.

\begin{table}[h!]
\centerline{TABLE 4}
\centerline{Stellar Photospheric and Dust Flux Density Contributions \label{tbl-4}}
\vspace{0.2cm}
\footnotesize
\begin{center}
\begin{tabular}{llcccc}
\tableline\tableline
Region  & Wavelength  & Total\tablenotemark{a}  & Dust & Stellar\tablenotemark{b}&Dust Fraction\\
        & \mum  & mJy  & mJy & &\\
\tableline
North   &3.6    &  20.8   &   0.0 & 1.0   & 0.00 \\
        &4.5    &  12.6   &   0.0 & 0.60  & 0.00 \\
        &5.8    &  15.2   &   6.4 & 0.42  & 0.42 \\
        &8.0    &  11.7   &   6.6 & 0.25  & 0.56 \\
        &14.3   &   6.9   &   5.2 & 0.08  & 0.75 \\
        &24.0   &  10.1   &   9.5 & 0.03  & 0.94 \\
        &70.0   & 104.2   & 104.2 & 0.00  & 1.00 \\
        &160.0  & 472.9   & 472.9 & 0.00  & 1.00 \\
\tableline
Center  &3.6    &  58.6   &   0.0 & 1.0  & 0.00 \\
        &4.5    &  35.4   &   0.0 & 0.60 & 0.00 \\
        &5.8    &  32.6   &   7.2 & 0.43 & 0.22 \\
        &8.0    &  25.5   &  10.7 & 0.25 & 0.42 \\
        &14.3   &  14.9   &   9.8 & 0.08 & 0.66 \\
        &24.0   &  17.6   &  15.7 & 0.03 & 0.89 \\
        &70.0   & 181.5   & 181.5 & 0.00 & 1.00 \\
        &160.0  & 712.9   & 712.9 & 0.00 & 1.00 \\
\tableline
South   &3.6    &   54.1   &    0.0& 1.0  & 0.00 \\
        &4.5    &   32.8   &    0.0& 0.60 & 0.00 \\
        &5.8    &   38.8   &   15.5& 0.40 & 0.40 \\
        &8.0    &   25.7   &   12.1& 0.23 & 0.47 \\
        &14.3   &   25.5   &   20.9& 0.09 & 0.82 \\
        &24.0   &   25.8   &   24.0& 0.03 & 0.93 \\
        &70.0   &  300.8   &  300.8& 0.00 & 1.00 \\
        &160.0  & 1490.0   & 1490.0& 0.00 & 1.00 \\
\tableline
Total   &3.6    &  437.6   &    0.0& 1.0  & 0.00 \\
        &4.5    &  265.7   &    0.0& 0.61 & 0.00 \\
        &5.8    &  279.0   &   89.3& 0.43 & 0.32 \\
        &8.0    &  147.9   &   37.0& 0.25 & 0.25 \\
        &14.3   &  101.4   &   63.9& 0.09 & 0.63 \\
        &24.0   &  141.4   &  127.2& 0.03 & 0.90 \\
        &70.0   & 1428.0   & 1428.0& 0.00 & 1.00 \\
        &160.0  & 8761.3   & 8761.3& 0.00 & 1.00 \\
\tableline
Core    &3.6    &  23.1   &   0.0 & 1.0  & 0.00 \\
        &4.5    &  14.9   &   0.0 & 0.64 & 0.00 \\
        &5.8    &  14.2   &   5.1 & 0.40 & 0.36 \\
        &8.0    &  14.5   &   9.3 & 0.22 & 0.64 \\
        &14.3   &   5.9   &   4.1 & 0.08 & 0.69 \\
        &24.0   &   5.1   &   4.4 & 0.03 & 0.87 \\
        &70.0   &  61.7   &  61.7 & 0.00 & 1.00 \\
        &160.0  & 253.2   & 253.2 & 0.00 & 1.00 \\
        &1100.0 &  21.0   &  21.0 & 0.00 & 1.00 \\
\tableline
\end{tabular}
\tablenotetext{a}{The measured flux density (photospheric plus dust contributions)}
\tablenotetext{b}{Stellar flux density normalized with respect to that at 3.6\mum}
\end{center}
\end{table}

\section{Mass of Dust and Gas in NGC~205}

With these new {\it Spitzer} observations, it is possible to derive
the mass of dust and estimate the gas mass in NGC~205 using different
methods and compare the results, for the total emission as well as
within the three main emission regions.

In the first method, we use a single-temperature fit to the 70 and
160\mum\ data to determine the dust temperature, assuming a modified
black body emissivity function of the form:

\begin{equation}
F(\lambda) = \lambda^{-\beta} \frac{2 h c^2 \lambda^{-5}}{e^{h c / k \lambda T} - 1}
\end{equation}

We adopt an emissivity coefficient $\beta=2$ (valid beyond 70\mum; cf.
Draine 2003). Assuming a gas-to-dust mass ratio of $\sim$100,
representative of the solar neighborhood and the inner disk of the
Milky Way (Hildebrand 1983; Devereux \& Young 1990), the dust
temperature and emissivity can be used to estimate the mass of dust
and gas.  This method has been extensively used to determine dust
masses in galaxies, but it has a few drawbacks (Devereux \& Young
1989; Dale \& Helou 2002) and should be taken as an approximation.
A fit to the spectral energy distribution (SED) has been shown to
provide better estimates to the dust masses for each dust grain
component (see below).

A reliable ratio of the far-IR fluxes is a necessary ingredient to get
the best temperature estimate. The angular resolution of the 70\mum\
image was reduced to match that at 160\mum\ by convolving the image
with the 160\mum\ kernel, derived from the PSF model generated using
the TinyTim software (Krist 2002).  The IRAF routines {\it geomap} and
{\it geotran} were then used to match the pixel scale and astrometry,
preserving the flux in the transformation.  Each pixel is assigned a
temperature based on the best fit to the modified blackbody
function. The resulting temperature map is shown in Figure~8. The
temperature is very uniform, ranging from $\sim 14 - 17$ K in the
innermost 5\arcmin\ that encompasses the three main emission regions.
In the central rectangular region of size 2.1\arcmin\ $\times$
2.9\arcmin\ (or an area of 6.3\arcmin$^2$, slightly larger than the
``Total'' region as defined in Table~1), we calculate the dust mass
from the 160\mum\ emission using the following equation:

\begin{equation}
M_d = S_{\nu} D^2 / \kappa_d B_{\nu,T}
\label{eq:mdust}
\end{equation}

\noindent where $D$ is the distance to the galaxy of 815~kpc and
$\kappa_d$ is the mass absorption coefficient for which we assume a
value of 6.73 m$^2$ kg$^{-1}$ at 70\mum\ and 1.2 m$^2$ kg$^{-1}$ at
160\mum\ (Li \& Draine 2001). The total dust mass in this region,
derived from the sum of the individual mass estimate for each pixel,
is $6.1\pm1.2 \times 10^4 M_\odot$, yielding a gas mass of $6.1 \times
10^6 M_\odot$ for a gas-to-dust mass ratio of 100.

Alternatively, we can determine dust masses by fitting the total dust
emission SED. The SED was measured between 5.8 and 160\mum, using
IRAC, ISO 14.3\mum\ and MIPS data. We measure the fluxes in the
spatially resolved regions common to all wavelengths, as depicted in
Figure~4 and listed in Table~3.

The dust emission in NGC~205 was modeled by an equation of the form

\begin{equation}
F_{dust}(\lambda) = \sum C_{i} \kappa_{i}(\lambda)
B_{\lambda}(T_{D,i})
\label{eq:dustfit}
\end{equation}

\noindent
where the the sum extends over $i$ dust components,
$C_{i}=M_{d,i}/D^{2}$, $D$ as in Eq. 2, and $\kappa_{i}$ is the mass
absorption coefficient of the i$^{th}$ dust component.  For NGC~205,
the relatively large 5.8, 8, and 14.3\mum\ fluxes require a hot
component which we ascribe to PAH molecules.  While the 70 and
160\mum\ emission could be fit with a single temperature component,
including the 24\mum\ and 1.1mm fluxes requires additional temperature
components. To satisfactorily fit the full SED of NGC~205, we adopt a
3 component model: warm and cold silicates ($a\sim0.1$\mum) and PAHs,
the former to reproduce the 24-1100\mum\ emission and the latter, the
5.8-24\mum\ emission. Mass absorption coefficients for astronomical
silicates were computed from Mie theory using the dielectric functions
of Laor \& Draine (1993). Cross sections for the PAH molecules were
taken from Li \& Draine (2001).  As the canonical PAH spectrum
(e.~g. NGC~7027, Werner et al. 2004) exhibits no features beyond
20\mum, we re-computed the PAH cross-sections (and mass absorption
coefficients) leaving off the last three terms of Li \& Draine's
Eq. 11.  The PAH component is included here for completeness of the
fit, but we do not use the temperature or mass estimates as the PAH
emission is a stochastic process rather than an equilibrium process as
assumed in Eq. 3.

\begin{figure}
\centerline{
\includegraphics[width=230pt,height=240pt,angle=-90]{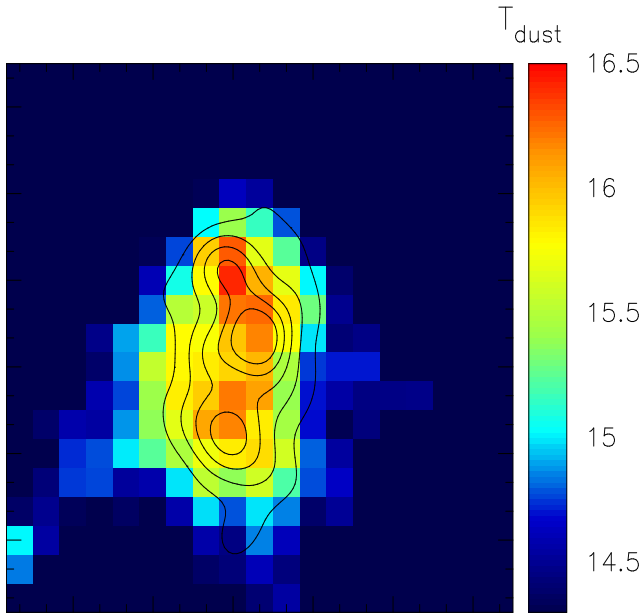}
}
\caption{\label{fig:8} Dust temperature of NGC~205 within the central
innermost 5\arcmin\ with 24\mum\ contours shown for levels 0.3, 0.35,
0.4, 0.45 and 0.5 MJy/sr. The temperatures are based on the 70 and
160\mum\ flux ratio. The 24 and 70\mum\ images have been convolved to
the 160\mum\ resolution.}
\end{figure}

With only 6 measurements and 6 parameters ($C_{1,2,3},T_{D_{1,2,3}}$),
a formal fit is not fully constrained and we proceed by making a
detailed search of parameter space. With our assumptions about the
components present in the observed SED, we searched a limited set of
parameters for dust mass (through $C_{i}$) and temperature for each
component.  We assumed temperature ranges of 200-550~K, 25-150~K, and
8-25~K for the PAH, warm, and cold silicates, respectively.  At each
set of temperature points, a similar grid of mass scalings ($C_{i}$)
was explored for each component.  After an initial, coarse grid
search, a finer temperature and mass scaling grid was defined for each
component and the procedure repeated.  We iterate this procedure
(typically only two refinements of the grid were needed) until a good
fit in the $\chi^2$ sense was found.  We estimated uncertainties in
the derived quantities by examining the distribution of $\chi^2$ about
the minimum for all input parameter sets (typically $1\times10^5 -
1\times10^6$ points in the six dimensional parameter space). While the
results of this procedure are not unique, it yields representative
estimates of both the dust mass and temperature, given the input
assumptions.  In all cases the dust mass is dominated by the cold
component and should be considered as a weak lower limit.  Owing to
the lack of data beyond 160\mum, we cannot formally rule out much
larger dust masses.  However, several pieces of evidence argue against
dust masses at the upper end of the formal error bars. First, where we
do have sub-mm data (``Core'' region), the dust mass is consistent
with much lower upper ``limits'' on the mass and second, the upper
error limits would imply very large fluxes at sub-mm wavelengths,
which is not what is observed (Fich \& Hodge 1991).

\begin{table}[h!]
\centerline{TABLE 5}
\centerline{Mass and Temperature Estimates from SED Fits \label{tbl-5}}
\vspace{0.2cm}
\begin{center}
\begin{tabular}{lcc}
\tableline\tableline
Cloud   &Warm Silicates   &Cold Silicates  \\
\tableline
North   &$1.2_{-1}^{+3} M_\odot$ & $2.2_{-1.5}^{+29} \times 10^{3} M_\odot$ \\
        &$53_{-5}^{+8}$~K          & $17.5_{-7}^{+5}$~K \\
\tableline
Center  &$2.1_{-2}^{+4} M_\odot$  &$2.9_{-1.7}^{+28} \times 10^{3} M_\odot$ \\
        & $53_{-6}^{+11}$~K          & $17_{-8}^{+3}$~K \\
\tableline
South   &$5.3_{-5}^{+4} M_\odot$  &$1.2_{-0.9}^{+2} \times 10^{4} M_\odot$ \\
        &$51_{-3}^{+18}$~K          &$15.8_{-3}^{+5}$~K \\
\tableline
Core\tablenotemark{a} &$0.6_{-0.3}^{+2.5} M_\odot$  &$1.1_{-0.9}^{+30} \times 10^{3} M_\odot$ \\
        &$53_{-8}^{+3}$~K          & $17.8_{-10}^{+5}$~K \\
Core\tablenotemark{b} &$1.9_{-1}^{+2} M_\odot$  &$1.8_{-0.4}^{+1.1} \times 10^{4} M_\odot$ \\
        &$48_{-4}^{+3}$~K          & $11.6_{-2}^{+3}$~K \\
\tableline
Total   &$6.1_{-5}^{+13} M_\odot$  &$3.2_{-1.8}^{+1.4} \times 10^{4} M_\odot$ \\
        &$58_{-5}^{+25}$~K          &$18.1_{-2}^{+2}$~K \\
\tableline
\end{tabular}
\tablenotetext{a}{SED without the 1.1mm flux density measurement}
\tablenotetext{b}{SED includes the 1.1mm flux density measurement}
\end{center}
\end{table}

The best SED fits to the data for each region, which produce the dust
masses and temperatures given in Table~5, are shown in Figure~9 and
10. Using the same gas-to-dust mass ratio of 100, we infer a gas mass
for the ``Total'' region of $3.2 \times 10^6 M_\odot$.  This value is
nearly one half of the $6.1 \times 10^6 M_\odot$ mass derived from the
70 to 160 flux ratio method.  Summing up the total dust mass in the
three distinct large regions using the best fit values, we estimate a
gas mass of $1.7 \times 10^6 M_\odot$, i.e.\ smaller than that of the
``Total'' region. This result suggests that a larger reservoir of
gas/dust with a comparable gas mass surrounds the three regions, as
first suggested by the HI (Young \& Lo 1997) and CO (Welch, Sage \&
Mitchell 1998) observations.

The ``Core'' region has a detection at 1.1mm (Fich \& Hodge 1991), and
adding this value to its SED and computing a new fit, results in a
dust mass of $1.8 \times 10^4 M_\odot$, i.e.\ approximately a factor
of sixteen larger than the mass estimated without the 1.1mm point
($1.1 \times 10^3 M_\odot$), and at a colder temperature, 11.6 K
(compared to 17.8 K). One can speculate that if this colder dust
component has a similar distribution across our selected regions,
preserving the same mass gas-to-dust ratio, then one predicts a gas
mass several times larger, $\sim 5\times 10^7 M_\odot$. This mass
determination is consistent with the fact that adding submillimeter
measurements to far-IR observations of early-type galaxies usually
yields larger cold dust mass estimates (as much as a factor ten) than
the far-IR measurements alone (Devereux \& Young 1990; Temi et
al. 2004).

\begin{subfigures}
\begin{figure}
\centerline{\includegraphics[width=260pt,height=200pt,angle=0]{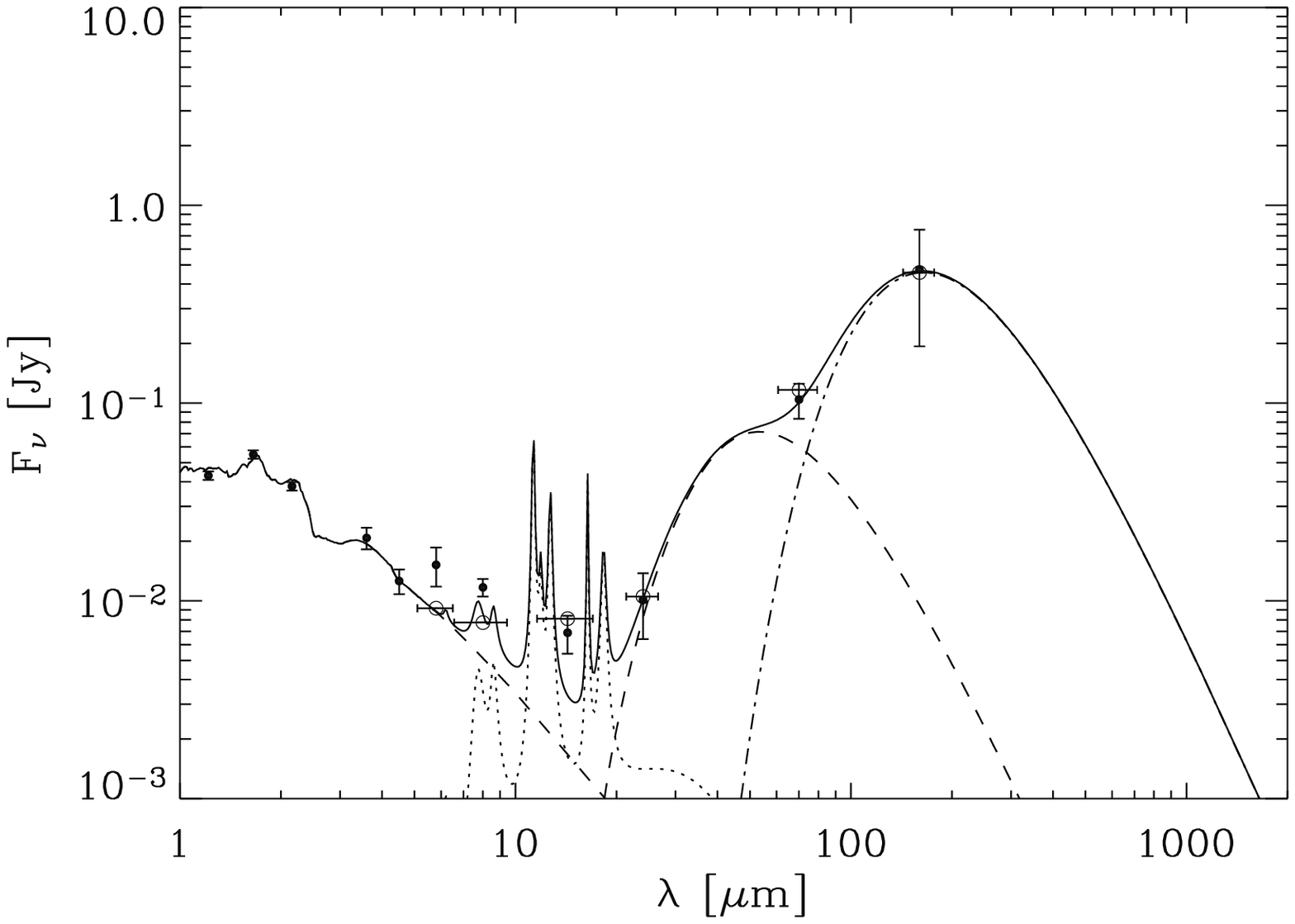}}
\centerline{\includegraphics[width=260pt,height=200pt,angle=0]{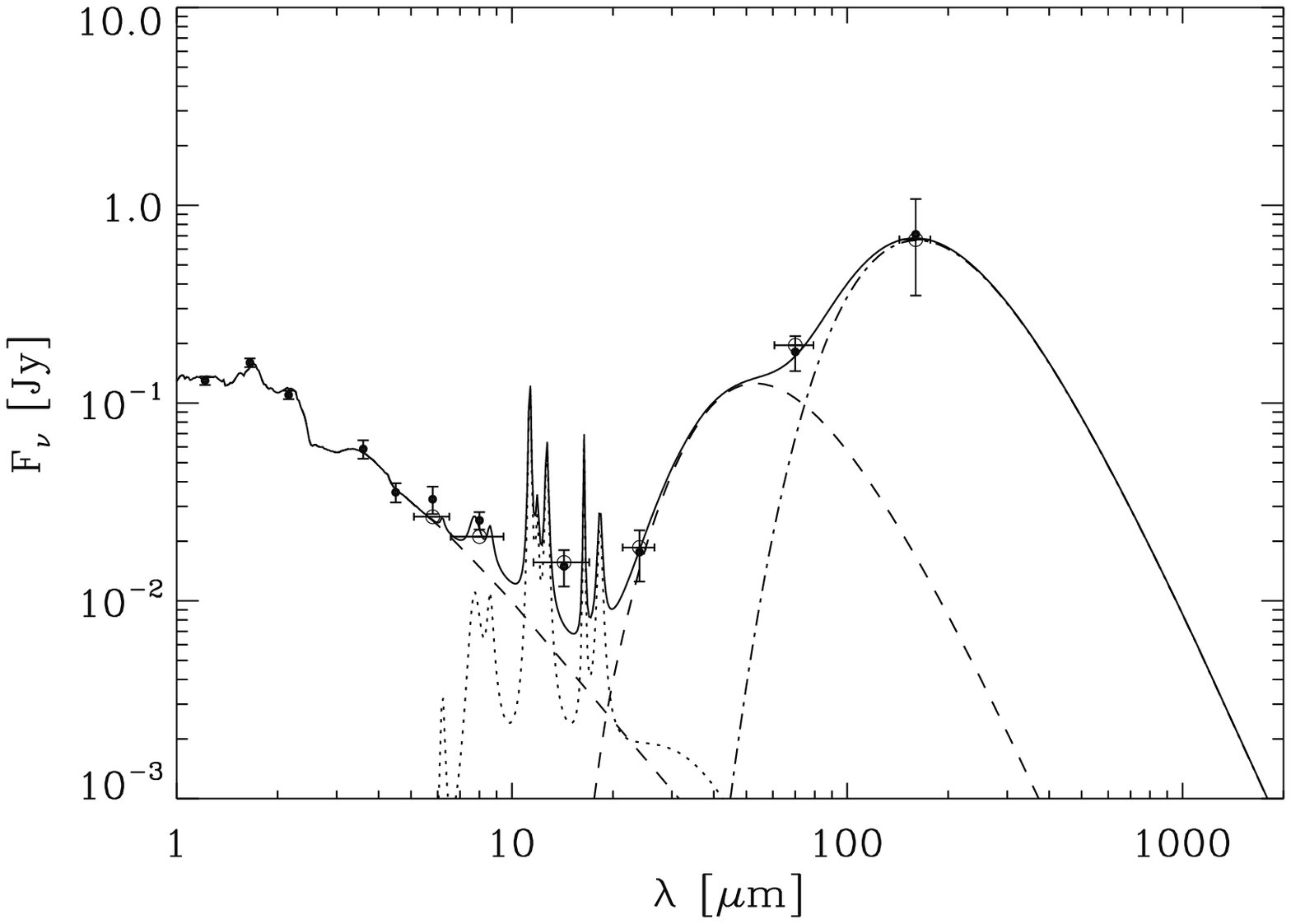}}
\caption{\label{fig:9a} Infrared SEDs of the
``North'' ({\it top left}) and ``Center'' ({\it top right}) region of NGC~205
(see Table 1). The diagrams show the SED fit to the IRAC, 14.3\mum\ ISO,
and MIPS measurements (Table 3). The {\it solid} points are the data
(stellar photospheric component removed) and the {\it open} points are
the model multiplied with the passband and integrated over wavelength.
The {\it dotted} line is the PAHs component, the {\it dashed} line is the
warm silicates component, the {\it dot-dashed} line is the cold silicates
component and the {\it solid} line is the total of all components.}
\end{figure}

\begin{figure}
\centerline{\includegraphics[width=260pt,height=200pt,angle=0]{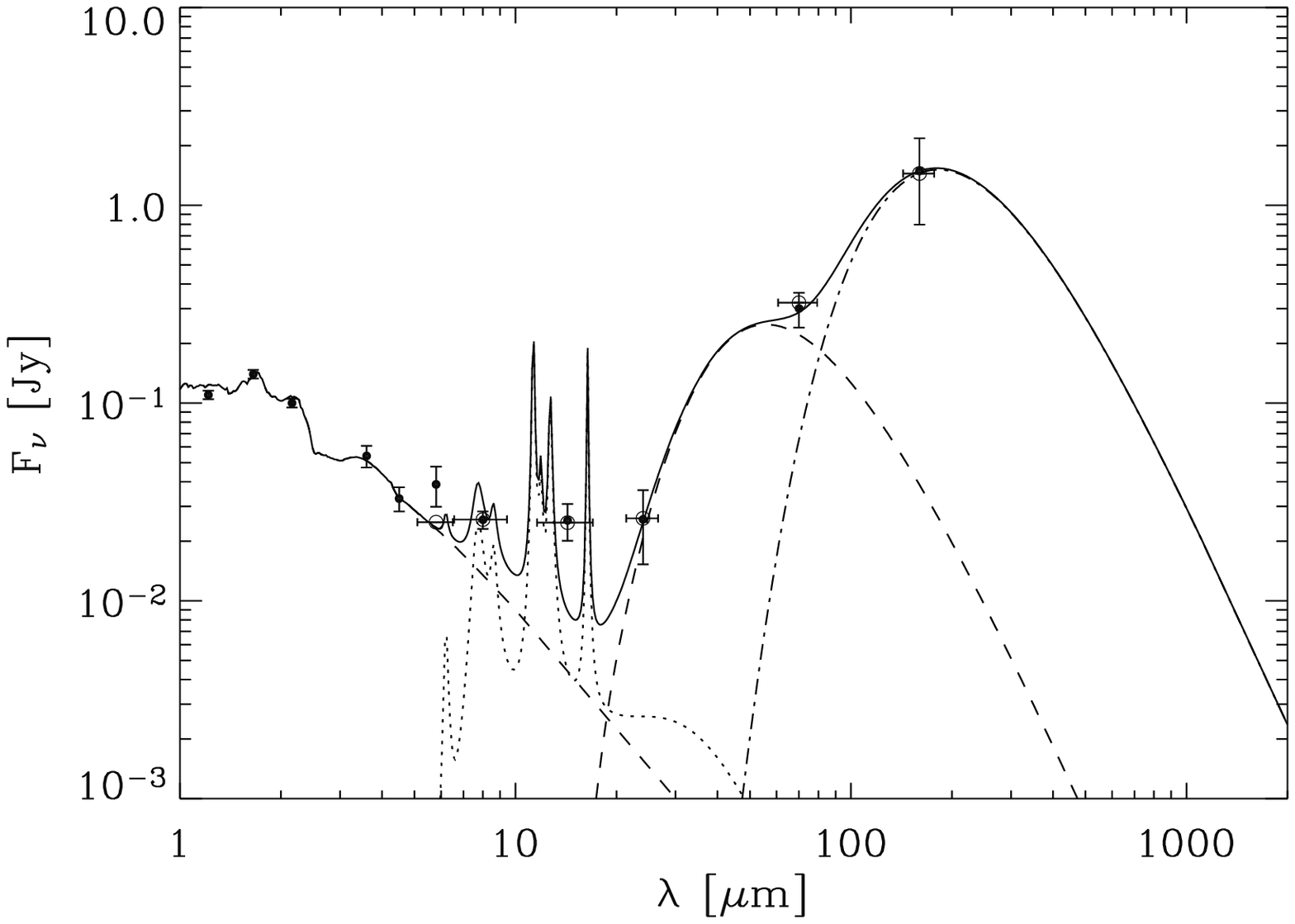}}
\centerline{\includegraphics[width=260pt,height=200pt,angle=0]{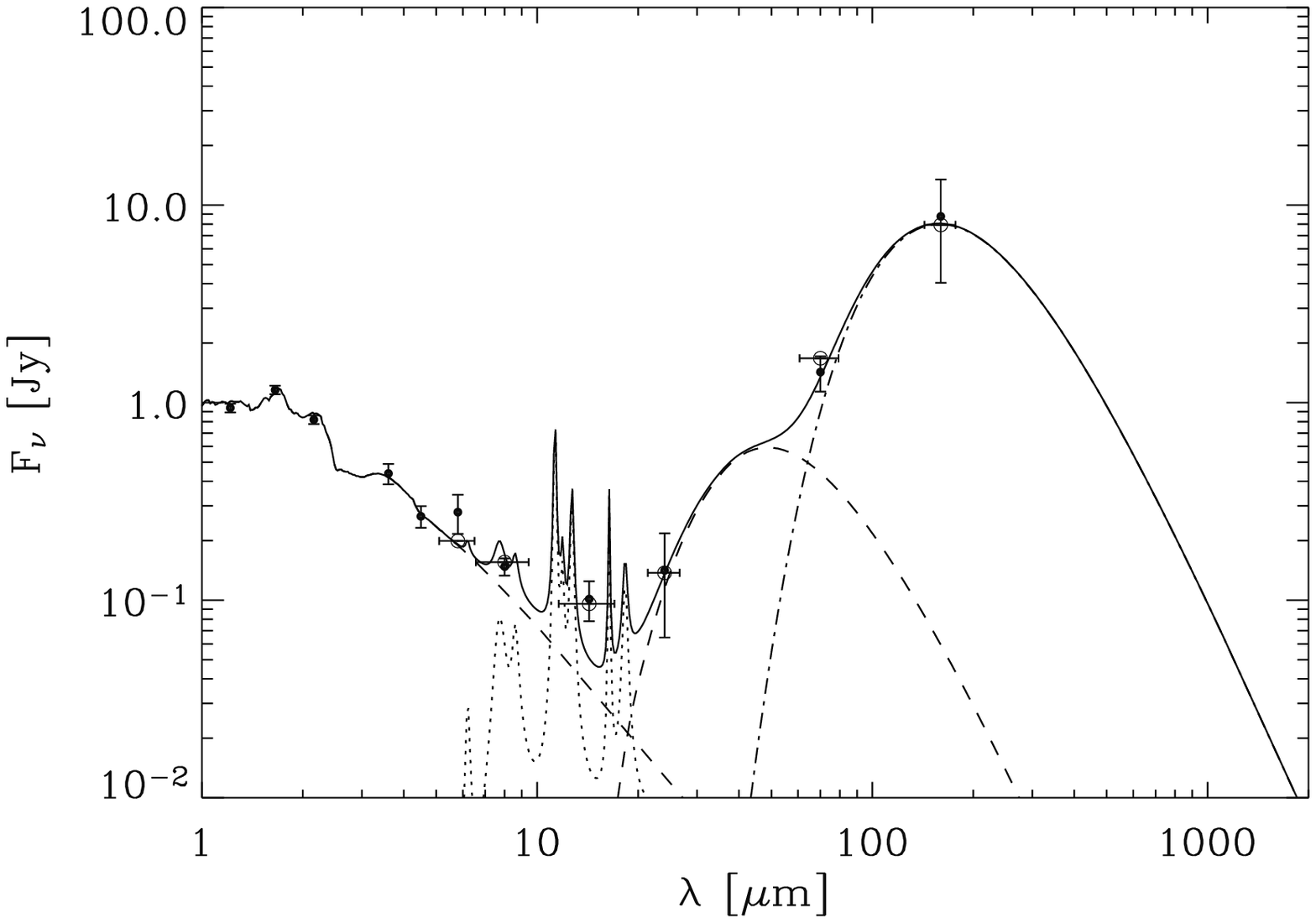}}
\caption{\label{fig:9ib} Infrared SEDs of the
``South'' ({\it bottom left}) and ``Total'' ({\it bottom right}) region of NGC~205 (see Table 1). }
\end{figure}
\end{subfigures}

\section{Discussion}

One of the open issues in understanding the ISM of NGC~205 has been
its ``missing ISM mass'', i.e. the apparent lack of gas in comparison
with the expected mass return to the interstellar medium by evolved
stars (Welch, Sage \& Mitchell 1998). By determining the dust mass and
selecting a reasonable gas-to-dust mass ratio (e.g.\ 100), one can
estimate the amount of gas in a straightforward manner.

The dust mass estimate for the ``Core'' of NGC~205 from Fich \& Hodge
(1991) using IRAS data and including 1.1mm observations is $M_{dust} =
1 - 3 \times 10^3 M_\odot$, depending on the adopted value for the
dust emissivity. The recent mass dust determination in the same region
by Haas (1998), adding ISO data between 120 and 200\mum\ to the SED,
is $M_{dust} = 0.88 \times 10^3 M_\odot$ {\it without} the 1.1mm data
and $M_{dust} = 1.18 \times 10^4 M_\odot$ including it (for a
emissivity exponent $\beta = 2$), i.e. a factor of four higher than
Fich \& Hodge's upper limit. Using the Spitzer data alone, we estimate
for the ``Core'' region $M_{dust} = 1.1 \times 10^3 M_\odot$. If we
include the 1.1mm data, the mass estimate increases to $M_{dust} = 1.8
\times 10^4 M_\odot$.  This latter value is similar to that obtained
by Haas, except that our dust temperature is 11.6 K rather than 8 K.
Therefore the mass of gas, for a gas-to-dust ratio of 100, for NGC~205
``Core'' region is $M_{gas} = 1.8 \times 10^6 M_\odot$.

Using only our FIR Spitzer data for the SED, we determine a ``Total''
amount of dust at 18 K of $M_{dust} = 3.2 \times 10^4 M_\odot$, which
implies a mass of gas, $M_{gas} = 3.2 \times 10^6 M_\odot$.  This
later value is four times larger than ISM mass determination of
$M_{ISM} = 7.2 \times 10^5 M_\odot$ by Sage, Welch \& Mitchell (1998),
and in perfect agreement with the theoretical gas mass estimate of
$M_{gas} = 2.6 \times 10^6 M_\odot$, generated by the injection of gas
by evolved stars over 0.5 Gyr.  This comparison suggests that the mass
determination by Sage, Welch \& Mitchell (1998) is indeed a lower
limit to ISM mass, and that essentially over our ``Total'' region one
can account for all the ISM expected in NGC~205 from the highly
evolved stars since the last starburst.

If the mass of dust in NGC~205 is indeed at a few times larger than
previous estimates, then we believe this provides a more coherent
picture of its recent star formation history. The global evolutionary
path of the ISM in NGC~205 was summarized by Welch, Sage \& Mitchell
(1998). They pointed out that any reasonable efficiency during the
last starburst in NGC~205 implies a more massive ISM. Their idea is
straightforward: if the burst transformed $\sim 1.4 \times 10^6
M_{\odot}$ of gas into stars (the Wilcots et al. 1990 value adjusted
to include low mass stars $ < 1 M_{\odot}$) with a 10\% efficiency,
then NGC~205 must have had at least $\sim 1.2 \times 10^7 M_{\odot}$
of gas. Our dust mass estimates could easily account for at least
$\sim 1/3$ of such a gas reservoir, and nearly all if the gas
temperature were $\sim 14$K.  Finally, a burst of $\sim 1.4 \times
10^6 M_{\odot}$ would yield $\sim 90$ stars in the brightest magnitude
of the AGB according to the fuel burning theorem of Renzini \& Buzzoni
(1986) [assuming a mass to light ratio of 0.2, the same as that of
NGC~1866, a $10^8$ year old star cluster (Fischer et al. 1992)].  This
is of the same order of magnitude as we see in Figure~7 (bottom).

\begin{figure}
\centerline{
\includegraphics[width=260pt,height=200pt,angle=0]{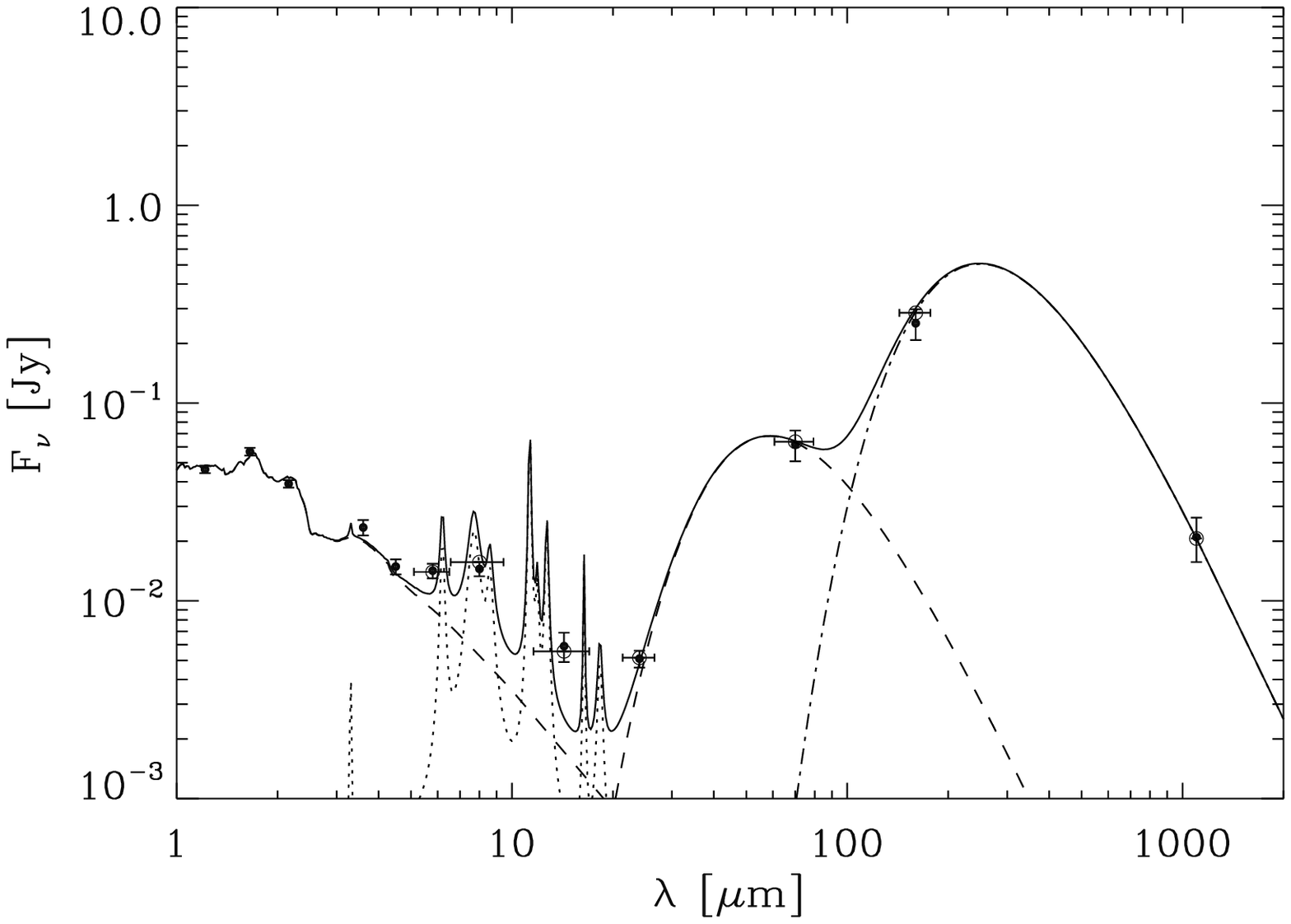}}
\caption{\label{fig:10} Infrared SEDs of the ``Core'' region of
NGC~205, as defined in Table~1. The diagram shows the SED
fit to the IRAC, 14.3\mum\ ISO, and MIPS measurements (Table 3)
plus the 1.1mm data point measured by Fich \& Hodge (1991).
The symbols and lines are defined in Fig.~9.}
\end{figure}

\section{Summary}

New IRAC and MIPS observations of NGC~205 give a better assessment of
the total dust mass in this prototypical LDG system. The {\it Spitzer}
data have been complemented with ISOCAM archival data to create a well
defined spectral energy distribution in the 3.6 to 160\mum\ wavelength
range.  The higher sensitivity and angular resolution of IRAC and MIPS
provide a detailed view of the morphology of the interstellar medium
in NGC~205. At 8 and 24\mum\ the spatial structure of the warm dust is
clearly resolved. Although the 8 and 24\mum\ images resemble the
drawings of the dust clouds inferred by Hodge (1973) using isodensity
measurements of the ground based visual photographic plates (c.f.\
Welch, Sage \& Mitchell 1998, Fig.~1; Young 2000, Fig.~1), these new
IR data display a more complex structure. To facilitate our analysis
of the color temperature and dust masses, we identified three main
regions.  As shown in Figure~2, the dust clouds within these regions
are bright and conspicuous at the longer wavelength too, having sizes
ranging from $\sim 100$ to 300~pc, and dust masses of $\sim 10^3 -
10^4 M_\odot$. A larger ``Total'' area encompasses the three regions,
and covers most of the bright 160\mum\ dust emission.

The derived dust masses, based on the 70 to 160\mum\ flux ratio and a
best fit to the SED, imply gas masses (assuming a standard gas-to-dust
mass ratio of 100 (cf Young 2000)), that are consistent with each
other and larger ($3.2-6.1 \times 10^6 M_\odot$) than previously
estimated from CO and HI observations.  This indicates that a
significant contribution to the dust and gas masses is coming from a
fainter component surrounding the brightest emission regions. Adding
the 1.1mm detection to our measurements of the ``Core'' region of
NGC~205, increases by a factor of sixteen the estimate of the cold
dust in that region. Our gas mass estimate in NGC~205 is a factor of
four larger than previously detected, but it could be still larger if
the distribution of the very cold dust ($\sim 12-14$ K) spreads out to
the three selected regions. If this is the case, the gas mass would be
sixteen times the ``Total'' (Table~5), i.e.\ $5 \times 10^7M_\odot$
for a gas-to-dust mass ratio of 100.  Overall the gas estimates are
therefore consistent with the predicted mass return from dying stars,
based on the last burst of star formation, $5 \times 10^8$ yr ago.

\acknowledgements

This work is based on observations made with the {\it Spitzer Space
Telescope}, which is operated by the Jet Propulsion Laboratory (JPL),
California Institute of Technology under NASA contract 1407.  Support
for this work was provided by NASA and through JPL Contract 1255094.
We thank the referee for her/his careful reading of the manuscript.

\end{document}